\documentclass{elsart1p}

\usepackage{amssymb}

\begin{document}

\begin{frontmatter}



\title{Gluon Propagator and Heavy Quark Potential in an Anisotropic QCD Plasma}


\author{Yun Guo}

\address{Helmholtz Research School,
Johann Wolfgang Goethe Universit\"at,
Max-von-Laue-Str.\ 1, D-60438 Frankfurt am Main, Germany\\
Institute of Particle Physics, Huazhong Normal University, Wuhan
430079, China}

\begin{abstract}
The hard-loop resummed propagator in an anisotropic QCD plasma in
general linear gauges are computed. We get the explicit
expressions of the gluon propagator in covariant gauge, Coulomb
gauge and temporal axial gauge. Considering one gluon exchange,
the potential between heavy quarks is defined through the Fourier
transform of the static propagator. We find that the potential
exhibits angular dependence and that there is stronger attraction
on distance scales on the order of the inverse Debye mass for
quark pairs aligned along the direction of anisotropy than for
transverse alignment.
\end{abstract}

\end{frontmatter}

\section{Introduction}
\label{intro}

Information on quarkonium spectral functions at high temperature
has started to emerge from lattice-QCD simulations; we refer to
ref.~\cite{Jakovac:2006sf} for recent work and for links to
earlier studies. This has motivated a number of attempts to
understand the lattice measurements within non-relativistic
potential models including finite temperature effects such as
screening~\cite{Mocsy:2005qw}. In this paper, we consider the effects due to a
local anisotropy of the plasma in momentum space on the
heavy-quark potential. Such deviations from perfect isotropy are
expected for a real plasma created in high-energy heavy-ion
collisions, which undergoes expansion. We derive the HTL
propagator of an anisotropic plasma for general linear gauges,
which allows us to define a non-relativistic potential via the
Fourier transform of the propagator in the static limit.

\section{Hard-Loop resummed gluon propagator in an anisotropic plasma}
\label{S-E and Propagtor}

The retarded gauge-field self-energy in the hard-loop
approximation is given by~\cite{ThomaMrow}
\begin{equation}
\Pi^{\mu \nu}(p)= g^2 \int \frac{d^3 {\bf k}}{(2\pi)^3} \,
v^{\mu} \frac{\partial f({\bf k})}{\partial k^\beta}
 \left( g^{\nu \beta} -
\frac{v^{\nu} p^\beta}{p\cdot v + i \epsilon}\right) \; .
\label{selfenergy1}
\end{equation}
Here, $v^{\mu} \equiv (1,{\bf k}/|{\bf k}|)$ is a light-like
vector describing the propagation of a plasma particle in
space-time. The self-energy is symmetric,
$\Pi^{\mu\nu}(p)=\Pi^{\nu\mu}(p)$, and transverse,
$p_\mu\Pi^{\mu\nu}(p)=0$.

In a suitable tensor basis the components of $\Pi^{\mu\nu}$ can be
determined explicitly. For anisotropic systems there are more
independent projectors than for the standard equilibrium
case~\cite{Romatschke:2003ms,Romatschke:2004jh,Mrowczynski:2004kv}.
We use a four-tensor basis developed in ref.~\cite{dgs:plb} and
the self-energy can now be written as $ \Pi^{\mu\nu}=\alpha
A^{\mu\nu}+\beta B^{\mu\nu} + \gamma C^{\mu\nu} + \delta
D^{\mu\nu}$ with

\begin{eqnarray}
\label{eq:basis}
A^{\mu \nu}&=& -g^{\mu\nu}+\frac{p^\mu
p^\nu}{p^2}+\frac{\tilde{m}^\mu \tilde{m}^\nu}{\tilde{m}^2},\,\,\,\,\,\
B^{\mu \nu}= -\frac{p^2}{(m\cdot p) ^2}\frac{\tilde{m}^\mu
\tilde{m}^\nu}{\tilde{m}^2} ,\nonumber\\
C^{\mu \nu}&=& \frac{\tilde{m}^2p^2}{\tilde{m}^2p^2+(n\cdot
p)^2}[\tilde{n}^\mu
\tilde{n}^\nu-\frac{\tilde{m}\cdot\tilde{n}}{\tilde{m}^2}(\tilde{m}^\mu
\tilde{n}^\nu+\tilde{m}^\nu
\tilde{n}^\mu)+\frac{(\tilde{m}\cdot\tilde{n})^2}{\tilde{m}^4}\tilde{m}^\mu
\tilde{m}^\nu],\nonumber\\
D^{\mu \nu}&=& \frac{p^2}{m\cdot p}
\left[ 2\frac{\tilde{m}\cdot\tilde{n}}{\tilde{m}^2}\tilde{m}^\mu
\tilde{m}^\nu-
\left(\tilde{n}^\mu \tilde{m}^\nu+\tilde{m}^\mu
\tilde{n}^\nu\right) \right]~.
\end{eqnarray}
Here, $m^{\mu}$ is the heat-bath vector, which in the local rest frame
is given by $m^{\mu}=(1,0,0,0)$, and $\tilde{m}^\mu=m^{\mu}-\frac{m\cdot p}{p^2} \,p^\mu$
is the part that is orthogonal to $p^\mu$.The direction of anisotropy in momentum space is determined by the
vector $n^{\mu}=(0,{\bf n})~,$ where ${\bf n}$ is a three-dimensional unit vector. We choose ${\bf n}=(0,0,1)$
in this paper. As before,
$\tilde{n}^\mu$ is the part of $n^\mu$ orthogonal to $p^\mu$.

In order to determine the four structure functions explicitly we employ
the following {\em ansatz}: $f({\bf p}) = f_{\rm iso}(\sqrt{{\bf p}^2+\xi({\bf p}\cdot{\bf
n})^2})$. The parameter $\xi$
is used to determine the degree of anisotropy. Thus, $f({\bf p})$
is obtained from an isotropic distribution $f_{\rm iso}(|\bf{p}|)$
by removing particles with a large momentum component along
$\bf{n}$. We do not list the rather cumbersome explicit
expressions for the four structure functions $\alpha$, $\beta$,
$\gamma$, and $\delta$ here since they have already been
determined in ref.~\cite{Romatschke:2003ms}.

The retarded propagator
$i\Delta^{\mu\nu}_{ab}$ is diagonal in color and so color
indices will be suppressed.  Using Dyson-Schwinger equation, its inverse is
given by
\begin{eqnarray}
&&\left(\Delta^{-1}\right)^{\mu \nu}(p,\xi)= -p^2 g^{\mu \nu}
+p^\mu p^\nu -\Pi^{\mu\nu}(p,\xi)+G~,
\end{eqnarray}
where $G$ is a gauge fixing term. In covariant gauge, Coulomb
gauge and temporal axial gauge, it has the following form
\begin{equation}
G_{cova}=-\frac{1}{\eta}p^{\mu}p^{\nu}\, ,G_{coul}=-\frac{1}{\eta}(p^{\mu}-\omega\,
 m^{\mu})(p^{\nu}-\omega\,
 m^{\nu})\, , G_{temp}=-\frac{1}{\eta}m^{\mu}
 m^{\nu}\,.
\label{inverse}
\end{equation}
Here, $\omega\equiv m \cdot p$ and $\eta$ is the gauge parameter. Upon inversion, the propagator in
covariant gauge is written as
\begin{equation}
\Delta^{\mu\nu}_{cova} = \frac{A^{\mu\nu} -
C^{\mu\nu}}{p^2-\alpha}  +
\Delta_{G}\left[(p^2-\alpha-\gamma)\frac{\omega^4}{p^4}B^{\mu\nu}
+ (\omega^2-\beta)C^{\mu\nu} +
\delta\frac{\omega^2}{p^2}D^{\mu\nu}\right] ~,
\end{equation}
where $\Delta^{-1}_{G} = (p^2-\alpha-\gamma)(\omega^2-\beta) - \delta^2
\left[{\bf{p}}^2-(n\cdot p)^2\right]$.

In Coulomb gauge, we have
\begin{equation}
\Delta^{\mu\nu}_{coul} = \frac{A^{\mu\nu} -
C^{\mu\nu}}{p^2-\alpha} +
\Delta_{G}\left[(p^2-\alpha-\gamma)\frac{\omega^2}{{\bf{p^2}}}\bar
{B}^{\mu\nu} + (\omega^2-\beta)C^{\mu\nu} +
\delta\frac{\omega^2}{p^2}\bar{D}^{\mu\nu}\right] ~,
\end{equation}
where the two new projectors $\bar{B}^{\mu\nu}$ and
$\bar{D}^{\mu\nu}$ are defined as
\begin{equation}
\bar{B}^{\mu\nu}= m^{\mu}m^{\nu}, \,\,
\bar{D}^{\mu\nu} =
\frac{p^2}{m\cdot p} \left[
2\frac{\tilde{m}\cdot\tilde{n}}{\tilde{m}^2}m^\mu m^\nu-
\left(\bar{n}^{*\mu} m^{\nu}+m^\mu \bar{n}^{*\nu}\right)
\right]~,
\end{equation}
with $\bar{n}^{*\mu} = n^{\mu}+\frac{n\cdot p}{{\bf{p^2}}}p^{\mu}$.

In temporal axial gauge, we have
\begin{equation}
\Delta^{\mu\nu}_{temp}  = \frac{A^{\mu\nu} -
C^{\mu\nu}}{p^2-\alpha} + \Delta_{G}\left[(p^2-\alpha-\gamma)\tilde
{B}^{\mu\nu} + (\omega^2-\beta)C^{\mu\nu} + \delta\tilde
{D}^{\mu\nu}\right]  ~,
\end{equation}
where the two new projectors $\tilde{B}^{\mu\nu}$ and
$\tilde{D}^{\mu\nu}$ are defined as
\begin{equation}
\tilde{B}^{\mu\nu}= \frac{(\omega\, m^{\mu}-p^{\mu})(\omega\, m^{\nu}-p^{\nu})}{{\bf{p^2}}} , \,\,
 \tilde{D}^{\mu\nu} =
(p^{\mu}-\omega\,
m^{\mu})\tilde{n}^{*\nu}+\tilde{n}^{*\mu}(p^{\nu}-\omega\, m^{\nu})~,
\end{equation}
with $\tilde{n}^{*\mu} = n^{\mu}-\frac{n\cdot
p}{{\bf{p^2}}}(\omega\, m^{\mu}-p^{\mu})$.

In the expressions of the propagators, we drop the gauge fixing
term. Actually, the gauge fixing term in covariant gauge, Coulomb
gauge and temporal axial gauge are $ -\frac{\eta}{p^4}p^\mu
p^\nu$, $ - \frac{\eta}{{\bf{p}}^4}p^\mu p^\nu$ and $-
\frac{\eta}{\omega^2}p^\mu p^\nu $, respectively.

It is easy to show that we recover the isotropic propagator by
setting $\xi=0$. In addition, if the gauge parameter $\eta=0$, we
can check in covariant gauge, $p_\mu \Delta^{\mu\nu}(p)=0$ because
of the gauge condition $\partial^\mu A_\mu=0$. In Coulomb gauge,
due to the fact that $\partial^i A_i=0$, we have $p_i \Delta^{\mu
i}(p)=0$ and in isotropic case, we have $\Delta^{0 i}(p)=0$. In
temporal axial gauge, we can check $\Delta^{0 i}=\Delta^{0 0}=0$
as a result of the gauge condition $A_0=0$.

\section{Heavy Quark Potential in an anisotropic plasma}
\label{potential}

We determine the real part of the heavy-quark potential in the
non-relativistic limit, at leading order, from the Fourier
transform of the static gluon propagator,
\begin{eqnarray}
V({\bf{r}},\xi) &=& -g^2 C_F\int \frac{d^3{\bf{p}}}{(2\pi)^3} \,
e^{i{\bf{p \cdot r}}}\Delta^{00}(\omega=0, \bf{p},\xi) \\
&=& -g^2 C_F\int \frac{d^3{\bf{p}}}{(2\pi)^3} \, e^{i{\bf{p \cdot
r}}} \frac{{\bf{p}}^2+m_\alpha^2+m_\gamma^2}
 {({\bf{p}}^2 + m_\alpha^2 +
     m_\gamma^2)({\bf{p}}^2+m_\beta^2)-m_\delta^4}~. \label{eq:FT_D00}
\end{eqnarray}
Here, $C_F$ is the color factor and the $\xi$-dependent masses
$m_\alpha^2$, $ m_\beta^2$, $m_\gamma^2$ and $ m_\delta^2$ are
given in ref.~\cite{dgs:plb}. This definition of potential should
be gauge independent. One can check that no matter the plasma is
isotropic or anisotropic, the definition is equivalent in
covariant gauge and Coulomb gauge. In temporal axial gauge, due to
the fact that $A_0=0$, it fails to define the potential. However,
there is a simple relation between these gauges: in static limit,
the quantity $\frac{\omega^2}{{\bf{p}}^2}\Delta^{i \,i}$ in
temporal axial gauge is identical to the quantity $\Delta^{00}$ in
covariant gauge and Coulomb gauge if the gauge fixing term
vanishes. In general, we find the quantity
$|\frac{\omega^2}{{\bf{p}}^2}\Delta^{i \,i}-\Delta^{00}|$ in
static limit is gauge independent which can be used as a more
general definition of the potential.

Generally, the integral
in~(\ref{eq:FT_D00}) has to be performed numerically. The poles of
the function are integrable. They are simple first-order poles
which can be evaluated using a principal part prescription. The numerical
results have been show in ref.~\cite{dgs:plb}. In
general, screening is reduced, i.e.\ that the potential at $\xi>0$
is deeper and closer to the vacuum potential than for an isotropic
medium. This is partly caused by the lower density of the
anisotropic plasma. However, the effect is not uniform in the
polar angle. Overall, one may therefore expect that quarkonium states whose
wave-functions are sensitive to the regime $\hat{r}\sim1$ are
bound more strongly in an anisotropic medium. Here, $\hat{r}\equiv rm_D$ and $m_D$ is the Debye mass.

\section{Conclusions}
\label{con}

We have determined the HTL gluon propagator in an anisotropic
plasma in general linear gauges. Its Fourier transform at
vanishing frequency defines a non-relativistic potential induced
by one gluon exchange for static sources. We find that,
generically, screening is weaker than in isotropic media and so
the potential is closer to that in vacuum. Also, there is stronger
binding of the quark pairs in the anisotropic system. Our results
are applicable when the momentum of the exchanged gluon is on the
order of the Debye mass $m_D$ or higher, i.e.\ for distances on
the order of $\lambda_D=1/m_D$ or less.  The binding energy for
quarkonium can be estimated analytically from this
potential~\cite{Guo:2008qm} if the quark mass is very large and
the temperature is very high. In this case we can neglect the
non-perturbative string contribution. For those states whose
length scale is larger, to determine the binding energy, we should
solve the Schr\"odinger equation with a potential which contains
the medium-dependent contributions due to one-gluon exchange {\em
and} due to the string~\cite{Mocsy:2007yj} (for a derivation of
the non-relativistic potential model see~\cite{brambilla}). This
is work in progress.

\section*{Acknowledgments}
The author gratefully acknowledges the collaboration with
A.~Dumitru and M.~Strickland and thanks SEWM 2008 for providing
the opportunity to present this work.


\begin{thebibliography}{00}

\bibitem{Jakovac:2006sf}
A.~Jakovac, P.~Petreczky, K.~Petrov and A.~Velytsky,
Phys.\ Rev.\  D {\bf 75}, (2007) 014506;
G.~Aarts, C.~Allton, M.~B.~Oktay, M.~Peardon and J.~I.~Skullerud,
Phys.\ Rev.\  D {\bf 76}, (2007) 094513.

\bibitem{Mocsy:2005qw}
A.~Mocsy and P.~Petreczky,
Phys.\ Rev.\  D {\bf 73}, (2006) 074007.

\bibitem{ThomaMrow}
S.~Mrowczynski and M.~H.~Thoma,
Phys.\ Rev.\  D {\bf 62}, (2000) 036011.

\bibitem{Romatschke:2003ms}
P.~Romatschke and M.~Strickland,
Phys.\ Rev.\ D {\bf 68}, (2003) 036004.

\bibitem{Romatschke:2004jh}
P.~Romatschke and M.~Strickland,
Phys.\ Rev.\  D {\bf 70}, (2004) 116006.

\bibitem{Mrowczynski:2004kv}
S.~Mrowczynski, A.~Rebhan and M.~Strickland,
Phys.\ Rev.\  D {\bf 70}, (2004) 025004.


\bibitem{dgs:plb}
A.~Dumitru, Y.~Guo and M.~Strickland,
Phys.\ Lett.\ B {\bf 662}, (2008) 37.

\bibitem{Guo:2008qm}
Y.~Guo,
arXiv:0805.2551 [hep-ph].

\bibitem{Mocsy:2007yj}
A.~Mocsy and P.~Petreczky,
Phys.\ Rev.\ D {\bf 77}, (2008) 014501.

\bibitem{brambilla}
N.~Brambilla, J.~Ghiglieri, A.~Vairo and P.~Petreczky,
Phys.\ Rev.\  D {\bf 78}, (2008) 014017.

\end{thebibliography}
\end{document}